\documentclass[twocolumn,showpacs,preprintnumbers,amsmath,amssymb,aps]{revtex4-1}
\usepackage{graphicx}
\usepackage{epsfig}
\usepackage{subfig}
\usepackage{dcolumn}
\usepackage{bm}
\usepackage{color}

\begin{document}

    \title{Topological Anderson insulating phases in the long-range Su-Schrieffer-Heeger model}
    \author{Hsiu-Chuan Hsu$^{1,2}$}\email{hcjhsu@nccu.edu.tw}
    \author{Tsung-Wei Chen$^3$}

    \affiliation{$^1$ Graduate Institute of Applied Physics, National Chengchi University, Taipei 11605, Taiwan\\
        $^2$ Department of Computer Science, National Chengchi University, Taipei 11605, Taiwan\\
    $^3$ Department of Physics, National Sun Yat-sen University, Kaohsiung 804, Taiwan}

    \date{\today}
\begin{abstract}
The long-range Su-Schrieffer-Heeger (SSH) model, in which the
second nearest-neighbor hopping is taken into account, exhibits a
topological phase diagram that contains winding numbers $w = 0,
1$, and $2$. In the clean system, the change in winding number stems from the band-touching phenomenon.  In the presence of disorder, the renormalization of energy band and Fermi level result in the nonzero density of states in the energy gap. These midgap states causes the crossover phenomenon and the divergence of localization length at a critical disorder strength $U_c$ in the finite SSH system. 
In this study, we numerically computed the
mean winding number and localization length for disordered SSH
system.  
We find that the disorder is able to drive phase
transitions between different mean winding numbers: $w$ = $0
\rightarrow 1$, $0 \rightarrow 2$, $1 \rightarrow 2$ and $2
\rightarrow 1$ in the weak disorder regime. By investigating the
wave function distribution and the self-energy,
the non-zero mean winding
numbers correspond to the so-called topological Anderson insulating (TAI)
phases. The finite size scaling for the mean winding number in the TAI phase is shown. For describing the phase transitions in the thermodynamic
limit, we apply the criterion of band gap closure resulting from
the broadening of energy band and Fermi level to determine the
critical disorder strength. The
critical disorder strength for self-consistent Born approximation
$U_c^{SCBA}$ and that for first Born approximation $U_c^{FBA}$ are
numerically calculated. $U^{SCBA}_c$ is found to match with $U_c$
qualitatively. 
Nonetheless, SCBA indicates the different roles of band shifts and Fermi level broadening near the topological phase transitions. Band shift / Fermi level broadening is more dominant for the transitions from low-to-high / high-to-low winding number.
Interestingly, for the transition from bulk insulator to TAI,
$U^{FBA}_c$ is quantitatively closer to $U_c$ than $U_c^{SCBA}$ as long as the renormalized band gap is zero within FBA.

\end{abstract}
\pacs{73.20.Fz, 73.21.Hb, 73.43.Nq, 73.63.Nm} \maketitle

\section{Introduction}
The phenomena of localization of electronic wave function in
random potential, now called Anderson localization, was proposed
by P. W. Anderson in his pioneering work in 1958
\cite{Anderson1958}. The scaling theory of localization shows that
in low dimensions, all states are localized no matter how weak the
disorder is\cite{Abrahams1979a,Abrahams2010}. As a result, in the
thermodynamic limit, any low dimensional system is an insulator.
In Ref.~\cite{Billy2008}, the Anderson localization is directly
observed  in one-dimensional matter waves of rubidium-87
Bose-Einstein condensates with controlled disorder. On the other
hand, when disorder is added to the system with energy band
topology, some interesting phenomena arise. It is known that
topological boundary modes are robust to disorder. In topological
insulators with high Fermi level that both bulk and boundary modes
transport, disorder plays a role that suppresses the bulk bands
and leaves the boundary modes conduct \cite{Du2016}. Furthermore,
in a normal insulator, disorder can drive the transition to a
topological insulator. This is called the topological Anderson
insulating (TAI) phase \cite{Li2009b}.

The TAI phase has been theoretically shown in several studies
\cite{Li2009b, Groth2009a, Guo2010a,Xu2012c,Altland2014,
Altland2015}. Li et al. \cite{Li2009b} showed that in a
two-dimensional BHZ model, disorder can lead to band inversion and
topological phase transition from normal to a nontrivial phase
that carries quantized conductance. In the same system, Groth et
al. \cite{Groth2009a} applied Born approximation to estimate the
renormalization of gap parameter that leads to inverted bands.
They conclude that the normal insulating and TAI phase boundary
correspond to the crossing of a band edge. Although the name seems
to suggest Anderson localization, the phase boundary actually
exhibits a weak disorder transition. Guo et al.\cite{Guo2010a}
also found that disorder transforms a normal insulator to a
topological insulator in three dimensions. Similar to the
conclusion drawn by Groth et al. \cite{Groth2009a}, the
weak-disorder boundary is the crossing of a band edge.
Nonetheless, they found the TAI phase extends to a regime where
energy broadening becomes significant and localization is the
leading factor. Xu et al. \cite{Xu2012c} showed there are two
kinds of TAI, the gapped and ungapped phases, in two-dimensional
BHZ model. In the gapped TAI, only edge states exist inside the
energy gap. In the ungapped TAI, the bulk and edge states coexist,
while the bulk states are localized by disorder. The latter is the
counterpart of the extended TAI in three dimension shown by Guo et
al. \cite{Guo2010a}.

On the other hand, Gergs et al. \cite{Gergs2016} showed that in
one-dimension Kitaev model, the topology is stabilized by
repulsive interaction and / or moderate disorder. Altland et al.
\cite{Altland2014, Altland2015} studied TAI phase transitions in
multichannel Su-Schrieffer-Heeger (SSH) chains and found
transitions between different winding numbers. They utilized field
theory within self-consistent Born approximation and two-parameter
renormalization group flow to locate the phase boundaries for
bulk/topological insulator (BI/TI) to TAI phases. It was shown in
their studies that disorder induces crossover to Anderson
insulator before the phase transition, which was determined by
delocalization and the half-integer winding number within SCBA. It
is only until recently that TAI phase has been observed
experimentally \cite{Meier929}. A one-dimensional SSH model that
preserves chiral symmetry was simulated in ultracold atoms. It was
shown that the winding number($w$) transitions from $w=0$ to $w=1$
as disorder strength increases. The experimental feasibility of
the SSH model \cite{Meier929,Xie2019,Fedorova2019} makes it
suitable for studying the interplay between band topology and
disorder.

Therefore, as motivated by these investigations, it is worthwhile
studying the rich phase diagram, the TAI phases and the scattering mechanisms in the SSH model. In this study, to explore the nontrivial phases with high
winding number in the presence of disorder, the long-range
interaction, which is experimentally applicable \cite{Xie2019}, is
included in the SSH model. TAI phases are shown in numerical
simulations and the mechanisms are explained with Born
approximation in the renormalized SSH system and theory of
localization in the finite SSH system. For the renormalized
system, the energy band is shifted and the Fermi level is
broadened by the imaginary parts of the self-energy. This enables
us to calculate the critical disorder strength by the closure of
band gap. In particular, we compare the phase boundary obtained
from the divergence of localization length with the band closure
within Born approximation. The crossover regions are observed in
our numerical results, which correspond to the nonzero imaginary
part of the self-energy in the renormalized SSH system. As
proposed in several theoretical works, the BI-TAI transition,
which corresponds to the $w=0$ to $w=1$ transition, is found.
Furthermore, we find that the transition can go directly from
$w=0$ to $w=2$ without crossing $w=1$. The TI-TAI is the
transition between two nontrivial insulating phases with different
winding numbers. The transitions $w=1\rightarrow 2\rightarrow 1$
and $w=2\rightarrow 1$ driven by disorder are found. 

This paper is organized as follows. In Sec.~\ref{sec:lrSSH}, the
model Hamiltonian and the methods for characterizing the TAI
phases are presented. The BI-TAI and TI-TAI transitions are
discussed in Sec.~\ref{sec:taiphase}. We also present the
thermodynamic limit for the fluctuation of winding number and conductance. In
Sec.~\ref{sec:born}, the crossover regions and band closure are
identified by self-energy within the Born approximation. The
critical disorder strengths are calculated by using Born
approximation and the comparison with the finite SSH system is
discussed. Our conclusion is given in Sec.~\ref{conclusion}.

\section{The long-range SSH model}\label{sec:lrSSH}
The one-dimensional Su-Schrieffer-Heeger (SSH) Hamiltonian with
long-range hopping that preserves chiral (sublattice) symmetry
\cite{Shem2014,Velasco2017,Maffei2018, Perez2019} is given by
    \begin{eqnarray}
    H_0&=&\sum_{i=0}^{N} J_0 C^{\dagger}_{i,a}C_{i,b} + J_1 C^{\dagger}_{i+1,a}C_{i,b} + J_2 C^{\dagger}_{i+2,a}C_{i,b}\nonumber\\
    &+&h.c.,
    \label{ham_x}
    \end{eqnarray}
where $i$ is the lattice site, $N$ is the length of the model,
$C^{\dagger}_{ia,ib}, C_{ia,ib}$ are the creation and annihilation
operators on sublattices $a, b$ on the $i$th unit cell. There are
two types of nearest neighbor coupling. $J_0$ denotes the
intracell coupling, while $J_1$ denotes the intercell coupling. In
the momentum space, the SSH Hamiltonian is written as
    \begin{eqnarray}
    H_0(k)&=&h_x(k)\sigma_x+h_y(k)\sigma_y,
    \label{ham_k}
    \end{eqnarray}
    where
    \begin{eqnarray}
    h_x(k)&=&J_0+J_1\cos k+J_2\cos(2k),\nonumber\\
    h_y(k)&=&J_1\sin k+J_2\sin(2k),
    \end{eqnarray}
$\sigma_{x,y}$ are Pauli matrices and act on the sublattices
$a,b$. The lattice constant is taken to be unity. The eigen energy
is $E_{\pm}=\pm\sqrt{h_x^2+h_y^2}$. Before directly calculating
the topological phase, the phase diagram can be inferred by
adiabatic connection \cite{Hsu2013a, Hsu2013b}.
    \begin{figure}[]
        \includegraphics[scale=0.5]{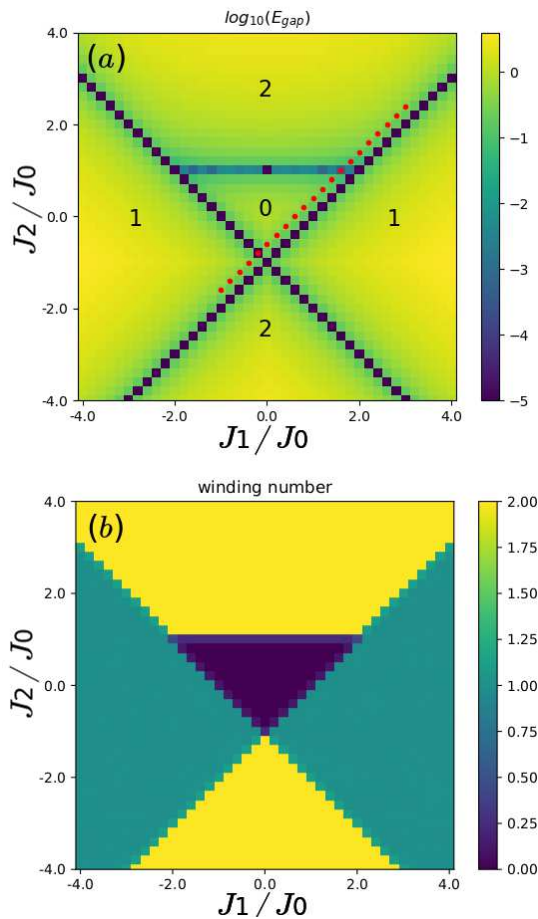}
        \caption{(color online) (a) The energy gap (b) The winding number as a function of $J_1, J_2$.
                 In (a), the winding numbers, annotated by the numbers on the plot, are inferred from
                 adiabatic connection. The red dashed line is in the vicinity of the phase boundary,
                 where the TAI phases are shown in Fig.\ref{TAIphasediagram}. In (b), the length of
                 chain is $Nx=400$ for winding numbers. Comparing the two figures, the vanishing energy
                 gap coincide with phase boundaries given by winding numbers.}
        \label{gapmap}
    \end{figure}

The energy gap as a function of $J_2/J_0$ and $J_1/J_0$ is shown
in Fig. \ref{gapmap}(a). The phase diagram is asymmetric about
$J_2=0$ due to the presence of the positive intracell term. The
gap closing conditions are given by
        \begin{eqnarray}
        &1&+\frac{J_1}{J_0}\cos k_0 +\frac{J_2}{J_0}\cos(2k_0)=0 \label{cond1},\\
        & &\frac{J_1}{J_0}\sin k_0 + \frac{J_2}{J_0} \sin(2k_0)=0 \label{cond2}.
        \end{eqnarray}
Eq. \ref{cond1} is satisfied when $k_0=0(\pi)$ and $J_0\pm
J_1+J_2=0$, giving rise to the straight phase boundary with slope
$\mp1$ and the interception at $J_2/J_0=-1$, as shown in
Fig.~\ref{gapmap}(a). Nonetheless, Eq. \ref{cond2} is also
satisfied by ${J_1}/{J_2}=-2\cos k_0$, which is plugged into Eq.
\ref{cond1} to obtain another condition ${J_2}/{J_0}=1$. As a
result, these conditions give rise to the horizontal boundary of
${J_2}/{J_0}=1$ for ${J_1}/{J_0}=[-2,2]$, as shown in
Fig.\ref{gapmap}(a).

The geometrical origin of the topology lies in the sublattice
pairing. In nontrivial topological phase, the bonding is formed
between opposite sublattices from different lattice sites, i.e.
sublattice $a$ bonds with sublattice $b$ at other site. This is
referred to as the singlet pairing \cite{Meier929}.
Quantitatively, this pairing is described by winding number
denoted as $w$ \cite{Shem2014, Meier929}. For $w=1$, the average
singlet pairing forms between the nearest neighbor, while for
$w=2$, the average singlet pairing forms between the next nearest
neighbor. In contrast, for trivial topology, the bonding is formed
within the same lattice site.

To study the effect of disorder on the topological phases, the
disordered intracell coupling is taken into account
    \begin{eqnarray}\label{disorderH}
    H_U=\sum_{i=1}^{N}U_iC^{\dagger}_{i,a}C_{i,b},
    \label{disorder}
    \end{eqnarray}
where $U_i$ are given by the random number in the range
$[\frac{-U}{2},\frac{U}{2}]$ with $U$ the disorder strengths in
the unit of $J_0$. For characterizing the topological phases in
disordered system, the winding numbers are computed numerically
for the tight-binding Hamiltonian.

Here, we use the method proposed by \cite{Shem2014, Meier929} that
applies for the chiral symmetric systems \cite{Maffei2018,
Meier929}.  By defining $Q=P_+-P_-$ and the chiral symmetry
operators $S=S_+-S_-$, where $P_{\pm}$ are the projection
operators that project to the positive or negative energy bands,
$S_{\pm}$ are the projection to sublattice a or b, the winding
number is given by
    \begin{eqnarray}
    w=-\mathrm{Tr_s}\left\{Q_{-+}[X,Q_{+-}]\right\},
    \label{wneq}
    \end{eqnarray}
where $Q_{+-}=S_+QS_-, Q_{-+}=S_-QS_+=(Q_{+-})^{-1}$, $X$ is the position operator and
$\mathrm{Tr_s}(\cdots)$ is the trace over the sublattices. This equation computes the local topological marker in real-space \cite{Bianco2011, Shem2014}. We calculate the average over the central part as the winding number for a chain \cite{Meier929}. For disordered systems, the average over disorder configurations is performed as well.  
The
winding number as a function of ${J_1}/{J_0}, {J_2}/{J_0}$ for the
clean limit is shown in Fig.~\ref{gapmap} (b). The results
coincide with that from adiabatic connection. However, the finite
size effect in the tight-binding Hamiltonian could smooths out the
phase boundaries. Therefore, the localization length, which peaks
identifies the topological phase transitions \cite{Shem2014} must
be also calculated.

To compute the localization length, the iterative Green's function
method is adopted \cite{Lewenkopf2013}. The localization length
can be extracted from the Green's function \cite{Mackinnon1983,
Kramer1993}
    \begin{eqnarray}
    \frac{2}{\lambda}=-\lim_{n\rightarrow \infty}\frac{1}{n}\mathrm{Tr}\ln |G_{1n}|^2,
    \end{eqnarray}
where $n$ is the total number of site of the SSH model, $G_{1n}$
is the propagator connecting the first and the last slice of the
system. A well-known challenge in this method is the vanishing
small eigenvalues due to successive matrix multiplication. To
overcome the numerical instability, we apply the method proposed
in \cite{Mackinnon1983} that normalizes the Green's function
regularly.

In the next section, we present the phase diagram driven by the
disorder defined in Eq. (\ref{disorderH}) along the trajectory of
the vicinity of the phase boundary [see Fig.~\ref{gapmap}(a)]. The
mean winding number is calculated and its fluctuation with the
length of the system is discussed.

\section{TAI phases and transitions}\label{sec:taiphase}

\begin{figure}[htb]
        \centering
    \includegraphics[scale=0.6]{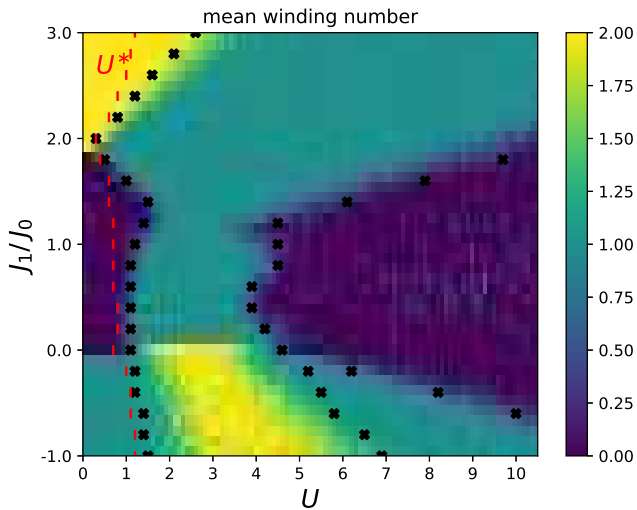}%
        \caption{(color online)
        The topological phases as a function of disorder strength and $J_1/J_0$ near the phase boundary, following the red dashed line
        in Fig. \ref{gapmap} (a). The black crosses denote the peaks of the localization length. The red dashed line denotes $U^{\ast}$, at which the
        self-energy starts to acquire an imaginary part (discussed in Sec.~\ref{sec:born}).}
        \label{TAIphasediagram}%
    \end{figure}

We study topological Anderson insulator (TAI) phase driven by the
disordered intracell coupling, as shown in Eq.~\ref{disorder}.
The TAI phase transition is investigated near the phase boundaries
in the clean limit. Fixing $J_0=1$ and $J_2=J_1-0.94$, the mean
winding numbers are plotted as a function of disorder strength and
$J_1/J_0$, as shown in Fig.\ref{TAIphasediagram}.

The winding number in the clean limit is retained up to a critical
disorder. As disorder strength increases, mean winding numbers
changes to another integer. In Fig. \ref{TAIphasediagram}, two
types of transitions are observed from this numerical result. The
transition from bulk insulator (BI) to TAI is the transition from
zero mean winding number to a non-zero mean winding number. Such
transition can be seen for $0\leq J_1/J_0 < 2$ in Fig.
\ref{TAIphasediagram}. The transition form topological insulator
(TI) to TAI is the transition between the two nontrivial
topological phases. It occurs for $-1\leq J_1/J_0<0$ and $2\leq J_1/J_0\leq 3$ at
weak disorder and $-1\leq J_1/J_0<0$ at strong disorder limit in Fig.
\ref{TAIphasediagram}.

\begin{figure}
    \includegraphics[width=0.43\textwidth]{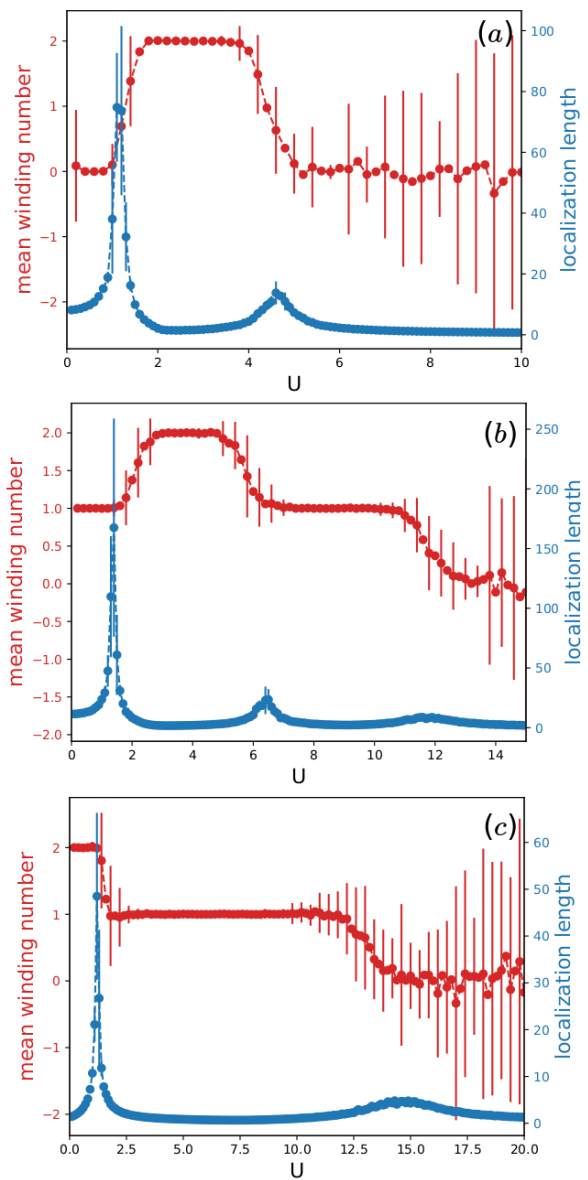}

        \caption{(color online) The mean winding number and localization length as a function of disorder strength for
        $(J_0, J_1, J_2)$. (a)$(1, 0, -0.94)$ (b) $(1, -0.8, -1.74)$ (c) $(1, 2.4, 1.46)$. For the mean winding number, the error bars are plotted
        every other data point for clarity. The number of disorder configuration for the mean winding number
and localization length is $50$ and $10$, respectively. The system
length for winding number is $1000$. The iteration steps for
calculating the localization length is $10^4$ that ensures
convergence.}
        \label{local}%
        \end{figure}

The transition would accompany the crossover phenomenon which will
be discussed in Sec.~\ref{sec:born}. Therefore, to better locate
the phase transitions, we compute the localization length. The
peaks of the localization length indicate the boundary of
topological phase transitions \cite{Shem2014,Habibi2018}. Our
numerical results show that the boundaries of the mean winding
number match the peaks of localization length as denoted by the
black crosses shown in Fig. \ref{TAIphasediagram}. The critical
disorder strength on the boundaries are denoted by $U_c$. Some
examples of the localization length and the changes in mean
winding numbers are given in Fig. \ref{local}. Fig. \ref{local}
(a) is a BI-TAI transition for $J_1=0$. Disorder drives the system
directly to $w=2$ because there is no singlet pairing between
nearest neighbor for this Hamiltonian.  When $J_1$ is nonzero,
disorder drives the formation of singlet pairing between the
nearest neighbor and the transition goes from $w=0$ to $w=1$. This
result shows that weak disorder scattering changes topological
properties of the system by strengthening the lowest-order nonzero
intercell coupling. Figs. \ref{local} (b) and (c) are TI-TAI
transitions. Disorder drives the transition between different
topological phases and eventually to normal insulating phase at
extremely strong disorder. The change of the mean winding number
is $1$, meaning only one pair of edge states is removed or formed
at a time. The second and third peaks are smaller, indicating that
the delocalization at transition is rather weak in the strong
disorder regime.

\begin{figure}[htbp]
	\centering
	\includegraphics[scale=0.5]{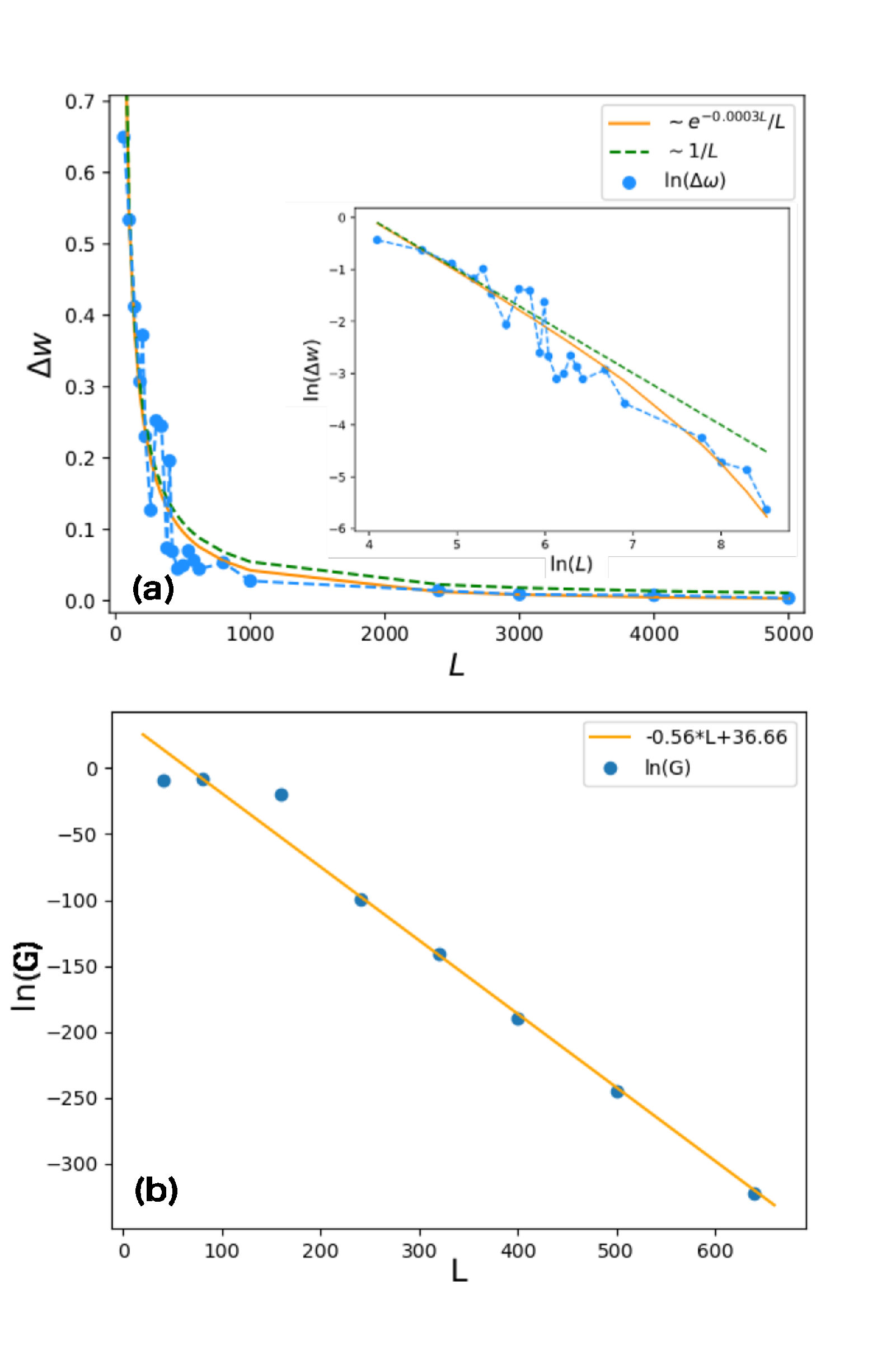}
	\caption{ (color online) (a) The scaling function of $\Delta w$ for $(J_0, J_1, J_2)=(1, 0, -0.94)$ and $U=2.5$. The inset shows the log-log plot. The dotted dashed line are the numerical values. (b) shows the scaling of the dimensionless conductance with chain length for the same parameters. The straight line  is the fitted linear relation between $\ln(G)$ v.s. $L$.}
	\label{scale}%
\end{figure}

The scaling functions of the fluctuation of the winding number, denoted as $\Delta w$, for TAI phase is studied. 
The exponential convergence of Eq. (\ref{wneq}) is shown by the rigorous mathematical proof in Ref.~\cite{Prodan2013}. 
Fig. \ref{scale}(a) shows that $\Delta w$ can be fit with $e^{-cL}/L $ with $c=0.0003$ (orange solid line in the figure).
 For the range of length considered, the exponential decay is not obvious, thus the decay is close to the algebraic $(L^{-1})$ decay, 
 as shown by the green dashed line in Fig. \ref{scale}(a).
The scaling shows that $\Delta w$ vanishes smoothly and $w$
reaches exact quantization in the thermodynamic limit
($L\rightarrow\infty$).
 Moreover, to further examine the
insulating phases, mean conductance as a function of chain length
was computed. The details about the computation of conductance is
described in the appendix. Fig. \ref{scale} (b) shows that the conductance
exponentially decays with chain length, i.e. $G \propto
e^{-L/\lambda}$,  confirming the insulating behavior in the TAI
phases. The numerical results agree with the
the two-parameter renormalization group theory \cite{Altland2014,
	Altland2015}, which states that both the mean conductance and the mean winding
number converge exponentially in the thermodynamic limit. 

\begin{figure*}%
            \centering
            \includegraphics[width=\textwidth]{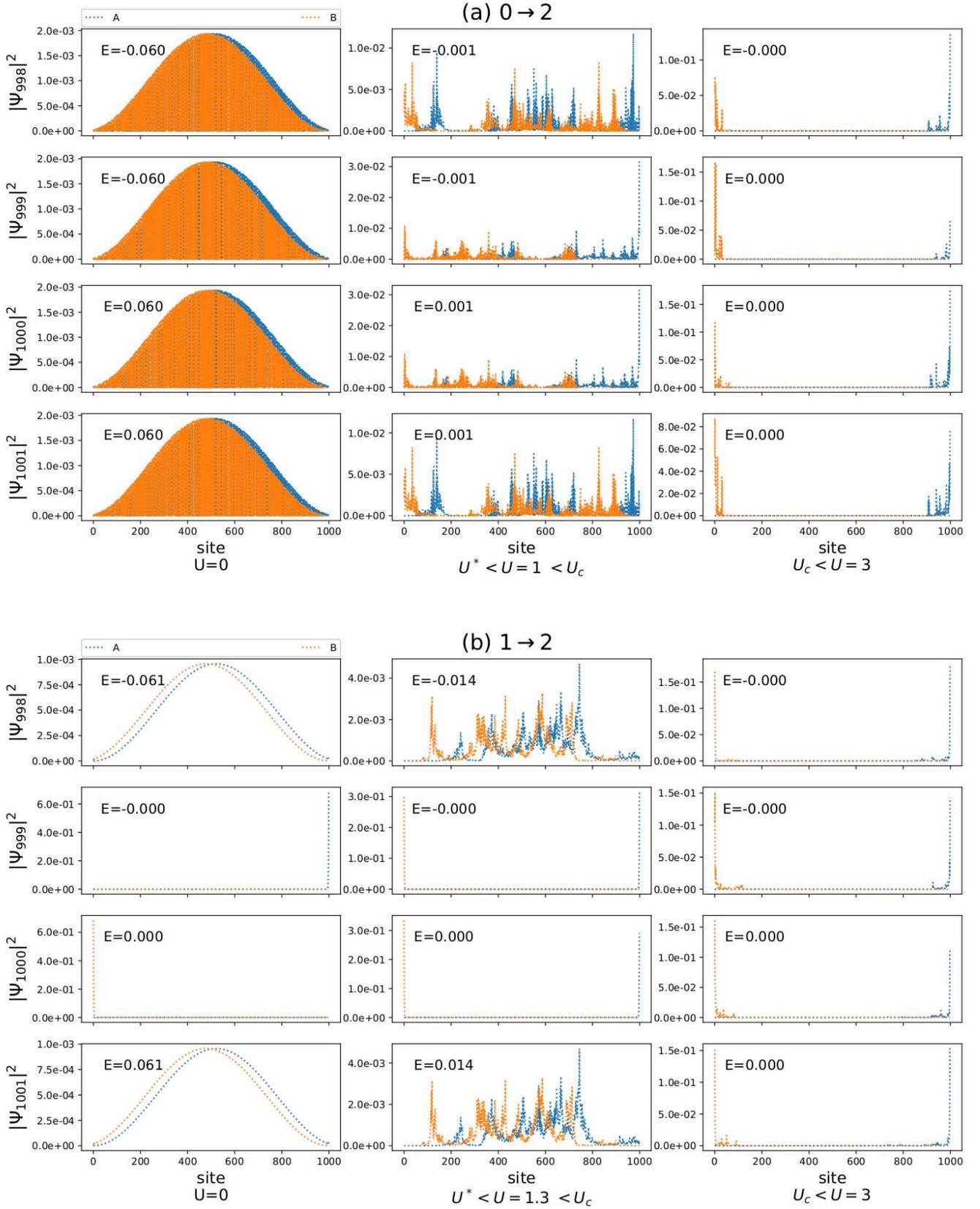}
            \caption{ (color online) Numerical results showing the probability density for sates near the band center in the clean limit (leftmost column),
            crossover (center column) and TAI phases (rightmost column) for $(J_0, J_1, J_2)$. (a) $(1, 0, -0.96)$ with $U^{\ast}=0.7$ and $U_c=1.1$, and (b)$(1, -0.8, -1.74)$ with $U^{\ast}=1.2$ and $U_c=1.4$.
            The number of disorder configuration is 20. The length of the chain is $1000$.}
            \label{dos}%
        \end{figure*}

The disorder drives
not only the phase transition but also the crossover phenomenon.
The latter can be seen by studying the probability distributions of
wave functions. Moreover, 
in topological phases, the bulk-edge correspondence predicts that
the numbers of pairs of edge states are the same as winding
numbers \cite{Kitaev2009,Chen2020}. We plot the disorder averaged probability density,
projected to each sublattice, of the states near the band center.
Figs. \ref{dos}(a) and (b) show the probability density near the
band center for TAI phases with $w=2$ via BI-TAI and TI-TAI
transitions, respectively. The probability densities were obtained
from direct diagonalization for a chain with length $1000$ lattice
sites. The $998^{th}$ to $1001^{th}$ states are plotted in each
row and the corresponding energy is shown in the legend.  The
disorder strength increases from the left to the right column. We
find that the center column is in the crossover regions for $
U^{\ast}< U <U_c$ and the rightmost column is in the TAI phase for
$U > U_c$.  The quantity $U^{\ast}$ is the disorder strength in
which the imaginary part of self-energy starts to be non-zero in
the renormalized SSH system. $U^{\ast}$ is labelled by the dashed
line (red) in Fig.~\ref{TAIphasediagram}. In the crossover
regions, the system transits from bulk insulator ($U=0$) to AI
($U^{\ast}<U<U_c$) and then becomes TAI ($U>U_c$). Fig.
\ref{dos}(a) shows the probability density along the
$w=0\rightarrow 2$ transition. The leftmost column ($U=0$) shows
the probability density in the clean limit, where the states near
the band center are bulk states and are away from zero-energy. The
rightmost column is the probability density at $U=3>U_c$, which is
in the TAI phase with $w=2$. There are two pairs of edge states at
$E=0$, the same as the winding number.

\begin{figure*}%
        \centering
        \includegraphics[width=\textwidth]{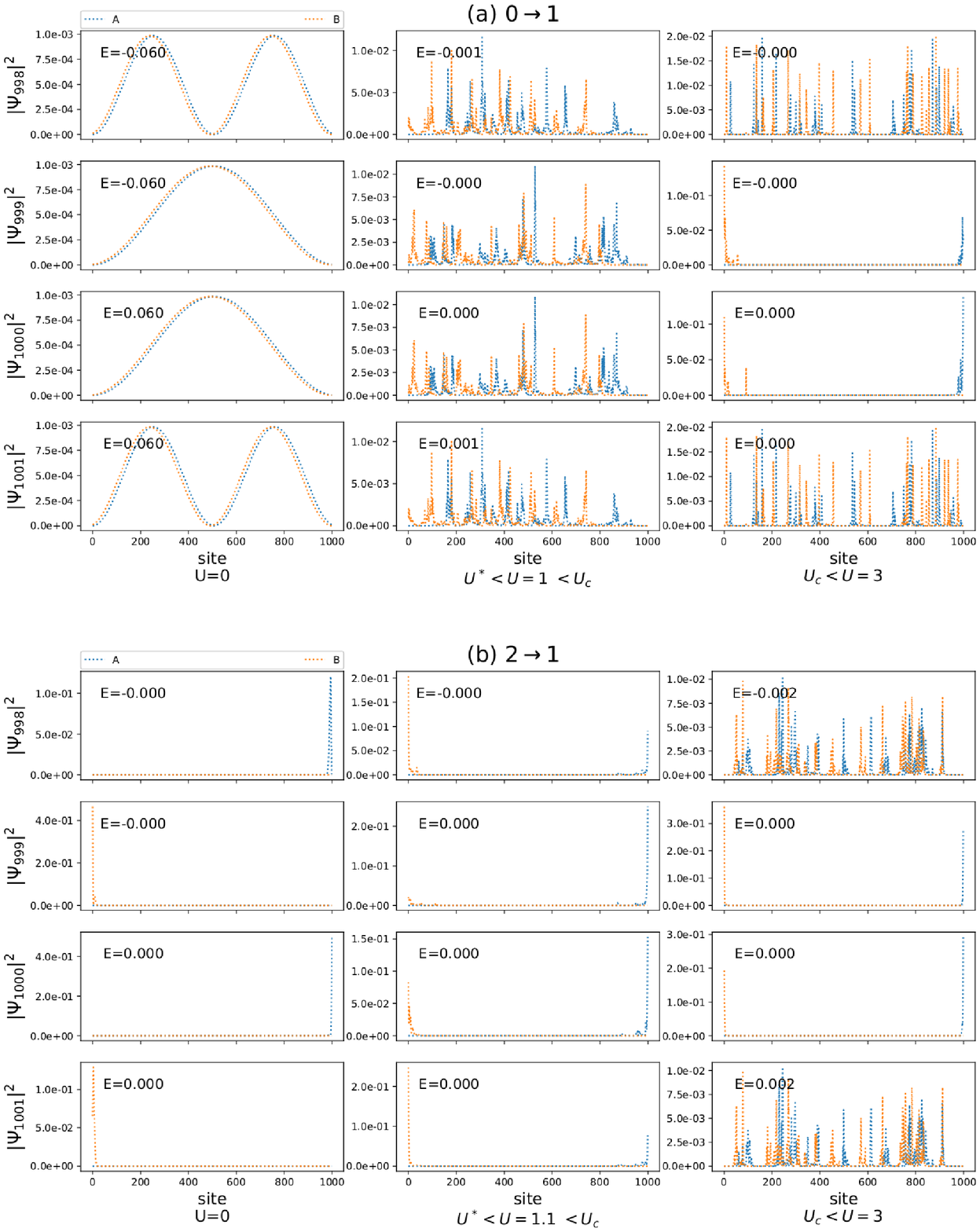}
        \caption{(color online) Numerical results showing the probability density for sates near the band center in the clean limit (leftmost column),
        crossover (center column) and TAI phases (rightmost column for $(J_0, J_1, J_2)$. (a) $(1, 1, 0.06)$ with $U^{\ast}=0.7$ and $U_c=1.2$, and (b) $(1, 2.4, 1.46)$ with $U^{\ast}=0.8$ and $U_c=1.2$. The number of disorder configuration is $20$. The length of the chain is $1000$.}
        \label{dosapp}%
    \end{figure*}

On the other hand, Fig. \ref{dos}(b) gives the probability density
along the $w=1\rightarrow 2$ transition (TI-TAI transition). There
is one pair of edge states at $E=0$ in the clean limit, as shown
in the leftmost column ($U=0$). In the crossover region
($U^{\ast}<U<U_c$), the bulk states becomes Anderson localized
states. Interestingly, The edge state is not significantly
affected by the disorder. The rightmost column is the probability
density at $U=3>U_c$, which is in the TAI phase with $w=2$. There
are two pairs of edge states at $E=0$, as expected by the
bulk-edge correspondence. The bulk-edge correspondence for TAI
phase with $w=1$ is also satisfied. The probability density along
the $w=0\rightarrow 1$ and $w=2\rightarrow 1$ transitions, are
presented in Fig. \ref{dosapp}. In Fig. \ref{dosapp}(b), the
winding number decreases from the clean limit to TAI phase. In the
crossover region, as shown in the center column, the probability
density gains few bulk contribution, indicating the robustness of edge states.
In the TAI phase with $w=1$, as shown in the
rightmost column, only one pair of edge state  is left at zero
energy, while the other pair is scattered into bulk with higher energy level. 

In short, to demonstrate the crossover, the probability density in
different regimes are plotted in Fig. \ref{dos}, \ref{dosapp}. In
Figs. \ref{dos}(a), (b) and Fig. \ref{dosapp}(a), the winding
number increases from the clean limit to TAI phase. The
probability density in crossover regions are shown in the center
column. The energy levels are drawn nearer to $E_f=0$ than that in
the clean limit. The spikes in the probability density demonstrate
the wave function localization in the bulk.

In the finite SSH system, it is shown that the phase transition
accompanies the crossover phenomenon. Nevertheless, in the
thermodynamic limit, the crossover phenomenon should appear when
the imaginary part of self-energy starts to be non-zero in the
renormalized SSH system. This is because the non-zero value of
imaginary part of self-energy would result in the broadening of
Fermi level (as well as energy band) and depicts more localized
states near the zero energy. In this sense, when the energy band
shift and the broadening of Fermi level together close the band
gap, the corresponding disorder strength for the closure of the
band gap would be the critical disorder strength. In
Sec.~\ref{sec:born}, we calculate the self energy and the critical
disorder strength by using Born approximation and compare the
results with the finite SSH system.

\section{Self energy and band closure}\label{sec:born}
As shown in Sec. \ref{sec:taiphase}, before the phase transitions,
there are crossover regions, where Anderson localization starts to
come into play\cite{Altland2014, Altland2015}. This is the
crossover regime when the BI (TI) enters the AI (TAI) phases
before phase transition. Within crossover, the topological edge
states do not significantly change while the density of states
penetrate into energy gap. The energy shift and Fermi level
broadening by disorder could close the band gap. In this sense,
the crossover phenomenon in the finite SSH system corresponds to
the non-zero imaginary part of self energy. Furthermore, the
closure of band gap can be used to determine the critical disorder
strength. We consider two approximations, First Born approximation
(FBA) and self-consistent Born approximation (SCBA). The
self-energy $\Sigma$ is given by the self-consistent equation
\begin{equation}\label{selfE}
\Sigma=\frac{U^2}{12}\sum_{k\in BZ}\frac{1}{z-H_0(k)-\Sigma},
\end{equation}
where $z=E_f+i\eta$. By regarding the self energy as
$\Sigma=\Sigma_x\sigma_x+\Sigma_0\sigma_0$ (which is consistent
with the numerical result), it can be shown that Eq. (\ref{selfE})
can be written as the renormalized $\bar{J_0}$ and $\bar{E}_f$
through the definitions $\bar{J_0}=J_0+\Sigma_x$ and
$\bar{E}_f=E_f-\Sigma_0$. We have
\begin{equation}\label{Eq-J0}
\bar{J}_0=J_0-\frac{U^2}{12}\sum_{k\in
BZ}\frac{\bar{J}_0+c_k}{(\bar{J}_0+c_k^2)^2+s_k^2-(\bar{E}_f+i\eta)^2},
\end{equation}
and
\begin{equation}\label{Eq-Ef}
\bar{E}_f=E_f+\frac{U^2}{12}\sum_{k\in
BZ}\frac{\bar{E}_f+i\eta}{(\bar{J}_0+c_k^2)^2+s_k^2-(\bar{E}_f+i\eta)^2},
\end{equation}
where $c_k=J_1\cos(k)+J_2\cos(2k)$ and
$s_k=J_1\sin(k)+J_2\sin(2k)$, where the lattice constant is taken
to be 1. The summation can be replaced by the integral, i.e.,
$\sum_{k\in BZ}=(1/2\pi)\int_{-\pi}^{\pi} dk$. The self-energy can
be solved analytically as given in Appendix.

\begin{figure}%
        \centering
         \includegraphics[width=0.45\textwidth]{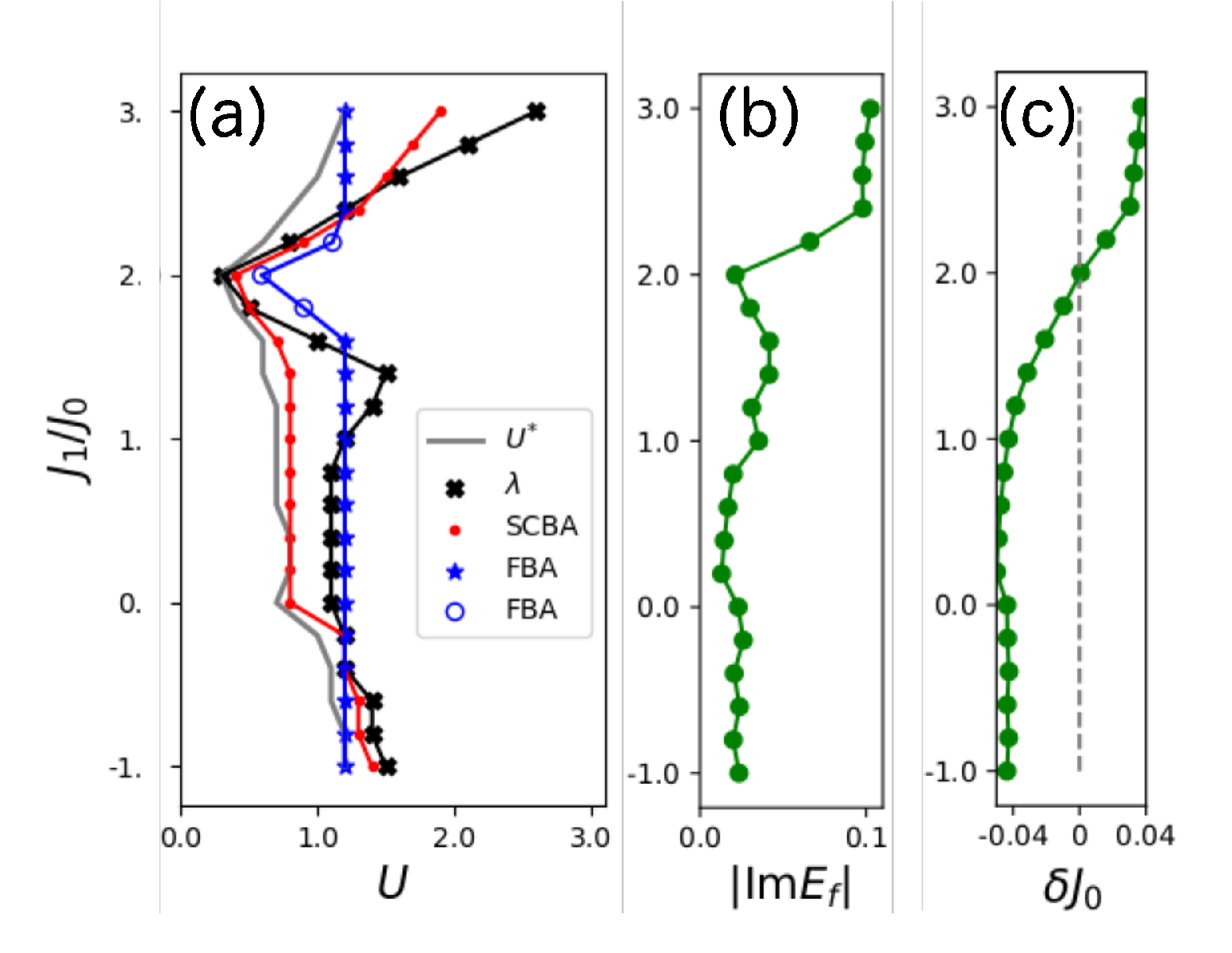} 
        \caption{ (color online) (a) Numerical results showing
the boundaries for band closure determined by SCBA (red $\bullet$)
and FBA (blue $\star$). The critical disorder strength $U_c$ is
shown by the cross (black $\times$). The region $-1\leq
J_1/J_0 < 0$ is the transition $w=1\rightarrow2$. $J_1/J_0=0$
is the transition $w=0\rightarrow2$. The region $0<J_1/J_0 <2$
is the transition $w=0\rightarrow1$. The region $2\leq
J_1/J_0\leq3$ is the transition $w=2\rightarrow1$. The empty circles along the FBA boundary is given by Eq. (\ref{condgap}) without satisfying Eq. (\ref{auxcond}). (b) The
imaginary part of the self-energy given by SCBA at band closure.
$\mathrm{Im}E_f$ is always negative. (c) $\delta J_0$ given by
SCBA at band closure.}
        \label{b}
    \end{figure}

In FBA, $\delta J_0=\bar{J}_0-J_0$ and $\delta E_f=\bar{E}_f-E_f$
are evaluated independently. At half-filling, i.e. $E_f=0$,
$\delta E_f$ is zero at weak disorder. The vanishing $\delta E_f$
implies that no crossover regions can be identified. 
The energy dispersion becomes $E=\pm\sqrt{(h_x+\delta J_0)^2+h_y^2}$. The condition for the gap closure is then given by
\begin{equation}
\begin{split}
&J_0+\delta J_0+J_1\cos k_0+J_2\cos(2k_0)=0,\\
&J_1\sin k_0+J_2\sin(2k_0)=0.
\end{split}\label{auxcond}
\end{equation}
For $k_0=\pi$, one obtains 
the critical
disorder strength in FBA denoted as $U_c^{FBA}$ by equating
\begin{eqnarray}
E_{min}=\frac{(U_c^{FBA})^2}{24J_0}, \label{condgap}
\end{eqnarray}
where $E_{min}$ is the lowest energy of the conduction band in the
clean limit. 
For $1.8 \leq J_1/J_0 \leq 2.2$, the band minima shift away from $\pi$,  to estimate $U_c^{FBA}$ for comparison with $U_c$, Eq. (\ref{condgap}) is applied even though Eq. (\ref{auxcond}) is not satisfied.
The critical disorder strength $U_c^{FBA}$ found by
this criterion is drawn on Fig. \ref{b} (a) with stars ($\star$) and open circles ($\circ$). We
find that $U^{FBA}_c$ exhibits approximately a straight line phase
boundary ($U^{FBA}_c\approx1.2$). For stronger $U_c>1.2$,
$U^{FBA}_c$ cannot qualitatively fit the results of $U_c$.
$E_{min}$ varies only in the regime $1.8 \leq J_1/J_0 \leq2.2$ and
is a constant beyond this regime, $U_c^{FBA}$ follows the same
trend as the bare values of $E_{min}$ according to
Eq.~(\ref{condgap}). The overall band closure boundary within FBA
is qualitatively different from the phase transition boundary.

In contrast to FBA, SCBA gives rise to the pure imaginary part
$\delta E_f$, which also affects $\delta J_0$ in the
self-consistent calculation. The numerical results show that the
self-energy has two components $\Sigma=\delta J_0\sigma_x + \delta
E_f\sigma_0$. Within SCBA, $\delta E_f$ acquires a nonzero
imaginary part, indicating the appearance of midgap states. The
corresponding disorder strength $(U^{\ast})$ is labelled by the
dashed line (red) in the phase diagram [see
Fig.\ref{TAIphasediagram}]. At band closure, the energy gap is
filled with electronic states when the band gap is renormalized by
$\delta J_0$ and smeared out by the Fermi level broadening in the
presence of disorder. Thus, the critical disorder strength
$U_c^{SCBA}$ is determined by
\begin{eqnarray}
\sqrt{(h_x(k_0)+\delta J_0)^2+h_y(k_0)^2}-|\delta E_f|=0.
\end{eqnarray}
and drawn on Fig. \ref{b}(a) with dots ($\bullet$). We find that
$U_c^{SCBA}$ gives a qualitative description of the phase
transition boundary ($U_c$).

In the renormalized SSH system, the change in the winding number
would be from the band shift and the localized state in the gap.
The former is determined by the value of $\delta J_0$ and the
latter by the value $\mathrm{Im}E_f$, which are shown in
Figs.~\ref{b}(b) and (c). From the numerical SCBA calculation, we
find that $\delta J_0$ changes sign. When $\delta J_0$ is negative,
the energy band minimum is pulled down, and the
bulk states at zero energy becomes significant. Our numerical results show that $U_c^{FBA}$ fits the trend with
$U_c$ in this regime. On the other hand, positive $\delta J_0$ leads to the rise
of the energy band minimum. In this case, our numerical results show
that $U^{SCBA}_c$ fits the trend with $U_c$ better than
$U^{FBA}_c$. For the case of vanishing $\delta J_0$, the change in
zero mode of edge states is due to the broadening of Fermi level,
and thus, $U^{SCBA}_c$ agrees with $U_c$. Furthermore, the
agreement between these values of $U_c$, $U^{SCBA}_c$ and
$U^{FBA}_c$ depends on the topological properties in the clean
limit. 
For systems in the topological states in the clean limit, $U^{SCBA}_c$ agrees with $U_c$ phase boundary. We discuss these numerical results in detail in the following.

In the region $0\leq J_1/J_0<1.8$ which is the BI-TAI transition
($0\rightarrow1$ and $0\rightarrow2$), FBA gives a quantitative
agreement to the phase boundary, as shown in Fig. \ref{b} (a). In this region, $\delta J_0$ is
negative and large, and $|\mathrm{Im}E_f|$ is small, indicating that the major contribution is from the band edge. At point
$J_1/J_0=1.8$, the transition is also from $w=0$ to $w=1$, i.e.,
BI-TAI transition, but $U^{SCBA}_c$ agrees with $U_c$ better than $U^{FBA}_c$ for this particular point. Since the renormalized band gap is not exactly zero within FBA, followed by the discussion along with Eq. (\ref{auxcond}, \ref{condgap}), $U_c^{FBA}$ does not agree with $U_c$.


In the region $-1\leq J_1/J_0<0$, which is the phase transition
$w=1\rightarrow2$. 
SCBA gives negative $\delta J_0$, suggesting that the extra zero mode edge states would be from bulk states. 
We observe that $U^{SCBA}_c$ is indeed
quantitatively closer to $U_c$ than $U^{FBA}_c$, especially in the
region with stronger $U_c$. Only in the region with weaker $U_c$,
we find that $U^{FBA}_c\approx U^{SCBA}_c\approx U_c$.
In contrast to the BI-TAI transition, there
exists zero mode edge states in the clean limit, $U^{SCBA}_c$
dominates the phase boundary, as a consequence of the zero mode
edge state being topologically protected up to all order of
scattering diagram.

On the other hand, for the transition from high to low winding
numbers ($2\rightarrow1$), i.e., the regime $J_1/J_0\geq2$, 
$U_c^{SCBA}$
fits the trend of the $U_c$ phase boundary better than
$U^{FBA}_c$, as shown in Fig.~\ref{b}(a). 
Fig.~\ref{b}(b, c) show that the
Fermi level broadening becomes significant. 
 This is in agreement with the probability density in finite system shown in the rightmost column in Fig. \ref{dosapp} (b). For reducing the winding number, one pair of the zero mode edge states is
 scattered into bulk.
 In contrast,  FBA still gives zero Fermi level broadening and negative band shift $\delta J_0$, i.e., the bulk state dominates the phase transition. 
As mentioned above, $U^{FBA}_c$ exhibits a straight line boundary
$U^{FBA}_c\approx1.2$.  When $J_1/J_0=2.4$, $U^{FBA}_c$ approaches
the straight line boundary, and numerically at this point we
obtain $U^{FBA}_c\approx U^{SCBA}_c$.

As pointed out by Guo et al. \cite{Guo2010a} in their study of the
three dimension TAI, the TAI regime where the self-energy obtains
an imaginary part is the "true" TAI phase. In three-dimensional
TAI, because bulk states do not contribute to conductance, bulk
states must be localized. In one dimension, we find that band
closure within SCBA gives a qualitative agreement with the phase
transition boundary. It suggests that the interplay between the
band edge renormalization and lifetime broadening are essential for
the phase transition. Similar to the three-dimensional TAI, we find that in
all TAI phases, the self-energy has a non-zero imaginary part, confirming the TAI phases. Nonetheless, we show that for high-to-low transition, the TAI phase has significantly larger lifetime broadening.

\section{Conclusion}\label{conclusion}

We have shown that the higher winding numbers exhibit in the
Su-Schrieffer-Heeger (SSH) system when the second nearest-neighbor
hopping is included in the SSH Hamiltonian. We study the
transition between bulk insulator (BI), topological insulator (TI)
and topological Anderson insulator (TAI). In the presence of
disorder, topological phase transitions driven by disorder are
identified by the divergence of localization length. The
disorder-induced phases are further investigated by the mean
winding numbers and wave functions. The scaling of the mean winding number is reported.  We calculated the critical disorder
strength by using first Born approximation (FBA) and
self-consistent Born approximation (SCBA). The critical disorder
strength for transitions are calculated by the criterion of the
closure of energy gap resulting from the broadening of energy band
and Fermi level. Compare to the the phase boundary given by delocalization ($U_c$), we showed that
the FBA exhibits phase boundary closer to $U_c$ than the SCBA for
BI-TAI transition if the renormalized band gap is zero within FBA. For TI-TAI
transition, we also showed the phase boundary obtained from SCBA
qualitatively fits $U_c$.
Moreover, SCBA shows that for the transition from low-to-high winding number, the band shift is more dominant, while for high-to-low winding number, the Fermi level broadening is more significant.

\acknowledgments 
We would like to thank D. W. Chiou for useful discussions on the winding number. 
The authors acknowledge the financial support by
the Ministry of Science and Technology of Taiwan through the
Grants MOST 108-2112-M-004-002-MY2 (H.C.H.) and MOST
108-2112-M-110-009 (T.W.C.).

\appendix
\section{Analytical calculation of Born Approximation}\label{sec:appa}
 Eq. (\ref{Eq-J0})
and (\ref{Eq-Ef}) can be evaluated analytically. The result of Eq.
(\ref{Eq-J0}) is given by (let $\eta\rightarrow0$)
\begin{equation}\label{Eq-J0-im}
\bar{J}_0=J_0-\frac{U^2}{12}\frac{1}{2\bar{J}_0}\left\{1+(-i)\frac{\sqrt{2}}{2}\left[\frac{1}{\Gamma_1}-\frac{1}{\Gamma_2}\right]\right\},
\end{equation}
\begin{widetext}
where
\begin{equation}
\begin{split}
&\frac{1}{\Gamma_1}=\frac{J_2\left[2\bar{J}_0^3+\bar{J}_0(2\bar{E}_f-J_1^2-2J_2^2)+J_1(J_1J_2+\sqrt{\mu_1})\right]}{\sqrt{\mu_1}\sqrt{M_1}},\\
&\frac{1}{\Gamma_2}=\frac{J_2\left[-2\bar{J}_0^3+\bar{J}_0(-2\bar{E}_f+J_1^2+2J_2^2)+J_1(-J_1J_2+\sqrt{\mu_1})\right]}{\sqrt{\mu_1}\sqrt{M_2}},
\end{split}
\end{equation}
and
\begin{equation}
\begin{split}
&\mu_1=-4\bar{J}_0^3J_2+J_1^2J_2^2-2\bar{J}_0J_2(-2\bar{E}_f+J_1^2+2J_2^2)+\bar{J}_0^2(J_1^2+8J_2^2),\\
&M_1=2\bar{J}_0^3J_2-\bar{J}_0^2(J_1^2-4J_2^2)-J_1J_2(J_1J_2+\sqrt{\mu_1})-\bar{J}_0(2\bar{E}_fJ_2-2J_2^3+J_1\sqrt{\mu_1}),\\
&M_2=2\bar{J}_0^3J_2-\bar{J}_0^2(J_1^2-4J_2^2)+J_1J_2(-J_1J_2+\sqrt{\mu_1})+\bar{J}_0(-2\bar{E}_fJ_2+2J_2^3+J_1\sqrt{\mu_1}).
\end{split}
\end{equation}
\end{widetext}
In the Born approximation (not self consistent), $\bar{J}_0$ and
$\bar{E}_f$ are replaced by the bare values $J_0$ and $E_f$ at the
right hand side of the equality of Eq. (\ref{Eq-J0-im}). In this
case, the second term in $\{\cdots\}$ of Eq. (\ref{Eq-J0-im}) is
pure imaginary and we have
\begin{equation}\label{Eq-J0-W}
\delta J_0\equiv-|E|=\Re[\bar{J}_0]-J_0=-\frac{U^2}{24J_0}.
\end{equation}
On the other hand, Eq. (\ref{Eq-Ef}) can also be evaluated
analytically, and the result is
\begin{equation}\label{Eq-Ef-1}
\bar{E}_f=E_f+(-i)\frac{U^2}{12}\bar{E}_f\sqrt{2}\left(\frac{1}{\Gamma_3}+\frac{1}{\Gamma_4}\right),
\end{equation}
where
\begin{equation}
\frac{1}{\Gamma_3}=\frac{\bar{J}_0J_2}{\sqrt{\mu_1}\sqrt{M_1}},~\frac{1}{\Gamma_4}=\frac{\bar{J}_0J_2}{\sqrt{\mu_1}\sqrt{M_2}}.
\end{equation}
In the Born approximation, $\bar{E}_f$ is replaced by $E_f$ at the
right hand side of Eq. (\ref{Eq-Ef-1}). Furthermore, in this
approximation, when $E_f=0$, we have $\bar{E}_f=0$.

\section{Calculation of the conductance}

To compute the length scaling of conductance, the
Landauer-Buttiker formula for two-terminal transport was utilized.
The conductance is related to the transmission function $(T)$ by
$(e^2/h)T$. The transmission function is given by \cite{dattabook}
\begin{eqnarray}
T=Tr[\Gamma_l G^R_{1L} \Gamma_r G^A_{1L} ]
\end{eqnarray}
where $G^{R(A)}_{1L}$ is the retarded (advanced) Green's function
corresponding to the transmission from the first to the last site
of the chain,  $\Gamma_{l (r)}$ is the surface self-energy of the
left (right) lead. The normal metal lead is attached to each end
of the SSH chain of various length (L). The dispersion of the
leads are taken to be $E=-2t_l cos(ka)$, where $t_l=2, a=1$ and
$k$ is the wave vector.

\begin{figure}
	\includegraphics[scale=0.5]{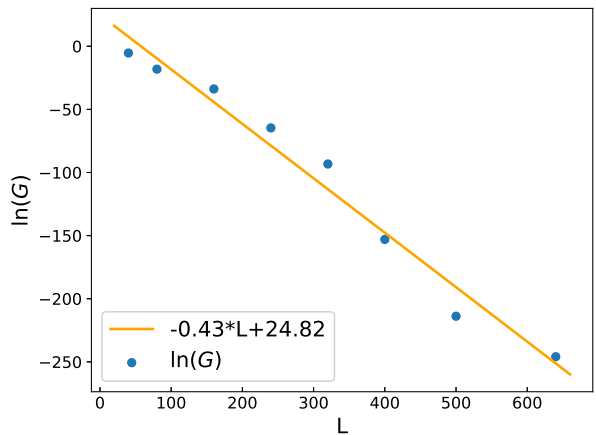}
	\caption{(color online) The scaling of the conductance with chain length for TAI phase for $(J_0,J_1, J_2)=(1, -0.8, -1,76)$ and $U=3.5$}
	\label{condscal}
\end{figure}

According to the theory of localization \cite{Rammer2004}, the
scaling function for the dimensionless conductance, denoted by
$G$,  is given by $\beta(G)={d\ln(G)}/{d\ln(L)}$. By chain rule,
\begin{eqnarray}
\beta(G)=\frac{d\ln(G)}{dL}\frac{dL}{d\ln(L)}=L\frac{d\ln(G)}{dL}.
\end{eqnarray}
In the limit of strong disorder, the conductance is assumed to be
proportional to $e^{-L/\lambda}$, where $L$ is the system
length and $\lambda$ is the localization length. Fig.
\ref{condscal} shows one example of the conductance scaling of TAI
regime (point (a) in Table I). The orange line in the plot is the
fitted linear relation between $\ln(G)$ and $L$. The scaling shows
that conductance exponentially decays with the system length and
confirms the insulating behavior in TAI phases.


\end{document}